\documentclass[prd, amsfonts, twocolumn, nofootinbib, showpacs]{revtex4}
\usepackage{graphicx, epsfig}
\usepackage{color}
\usepackage{amsmath}
\newcommand{\be}{\begin{equation}}
\newcommand{\ee}{\end{equation}}
\newcommand{\bea}{\begin{eqnarray}}
\newcommand{\eea}{\end{eqnarray}}

\newcommand{\gapp}{\mathrel{\raise.3ex\hbox{$>$}\mkern-14mu \lower0.6ex\hbox{$\sim$}}}
\newcommand{\lapp}{\mathrel{\raise.3ex\hbox{$<$}\mkern-14mu \lower0.6ex\hbox{$\sim$}}}
\def\bbox{{\,\lower0.9pt\vbox{\hrule \hbox{\vrule height 0.2 cm
\hskip 0.2 cm \vrule  height 0.2 cm}\hrule}\,}}

\begin{document}
\title{Using quasars as standard clocks for measuring cosmological redshift}
\author{De-Chang Dai$^1$, Glenn D. Starkman$^2$, Branislav Stojkovic$^3$, Dejan Stojkovic$^4$, Amanda Weltman$^1$}
\affiliation{$^1$ Astrophysics, Cosmology and Gravity Centre, University of Cape Town, Rondebosch, Private Bag, 7700, South Africa}
\affiliation{$^2$ CERCA/ISO/Department of Physics, Case Western Reserve University, Cleveland, OH 44106-7079}
\affiliation{$^3$ Department of Computer Science and Engineering, SUNY at Buffalo, Buffalo, NY 14260-1500}
\affiliation{$^4$ HEPCOS, Department of Physics, SUNY at Buffalo, Buffalo, NY 14260-1500}

\begin{abstract}
\widetext
We report hitherto unnoticed patterns in quasar light curves. We characterize segments of quasars' light curves with the slopes of the straight lines fit through them. These slopes appear to be directly related to the quasars' redshifts. Alternatively, using only global shifts in time and flux, we are able to find significant overlaps between the light curves of different pairs of quasars by fitting the ratio of their redshifts. We are then able to reliably determine the redshift of one quasar from another. This implies that one can use quasars as standard clocks, as we explicitly demonstrate by constructing two independent methods of finding the redshift of a quasar from its light curve.
\end{abstract}


\maketitle
{\bf Introduction.~}
To probe the largest distances in our universe, we seek the brightest  sources. Quasars are certainly among these, and they are numerous. Unfortunately quasars exhibit large dispersion in luminosities at all wavelengths. This makes them unusable as standard candles for measuring cosmological distances. In \cite{Baldwin}, an interesting correlation between emission line equivalent width (the ratio of integrated line flux over local continuum flux density) and rest-frame ultraviolet luminosity was observed. Namely,  C IV 1549 A emission-line equivalent width in quasars decreases with increasing UV continuum (1450 A) luminosity. Since flux ratios are distance-independent, this anti-correlation could allow for quasars to be used as standard candles. However, the large dispersion in this anti-correlation gives very poor distance calibrations compared to other standard candles. It is therefore worthwhile to look for other correlations that might make quasars viable distance standards.

{\bf Identifying the patterns in quasars' light curves.~}
In this paper, we analyze the V-band light curve data from MACHO spectroscopically-confirmed quasars behind the Magellanic Clouds \cite{Geha:2002gv}.
These light curves and spectra  are available at
$http://www.astro.yale.edu/mgeha/MACHO/$.
Most analyses of quasar light curves have considered  the variation of their absolute magnitude \cite{Hawkins:2010xg,Hawkins:2001be,Pelt,Wold:2006eb},
whereas we will  use  the flux, which is more physical. The relation is
\begin{equation}
m = -2.5 \log_{10} (F/F^0)\,.
\end{equation}
Here $F$ is the flux,
$F^0$ is a V-band-specific normalization constant ($3.636\times 10^{-20} ergs/cm^{2}/sec/hz$), and $m$ is the quasar's absolute V-band magnitude.

\begin{figure}[h]
   \centering
\includegraphics[width=3.2in]{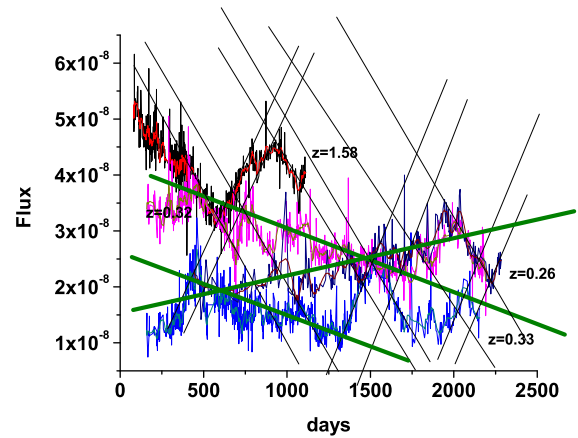}
\caption{The sample of quasars with different redshifts. Time  is transformed to the quasar rest frame (i.e. divided by $(z+1)$). A pattern emerges, some parallel lines (same slopes) appear. We mark them with the thin black lines. A possible shallower, long-term trend is indicated by thick green lines. }
    \label{normalize}
\end{figure}

To study intrinsic patterns of quasar activity it is best to transform to the quasar's rest frame, i.e. at minimum we need to rescale the observed time  by $(1+z)^{-1}$.
In Fig.~\ref{normalize}, we plot the flux, $F$, as a function of quasar-rest-frame time for selected quasars
from among the MACHO sample.
Two distinct  patterns of time variation are apparent as linear trends in these rest-frame light curves.
To emphasize these trends we mark them with parallel lines of two distinct slopes.
Shorter time-scale trends (marked by the steeper thin black lines) are superimposed on the longer time-scale trends
(thick green lines).

\begin{figure}[h]
   \centering
\includegraphics[width=3.2in]{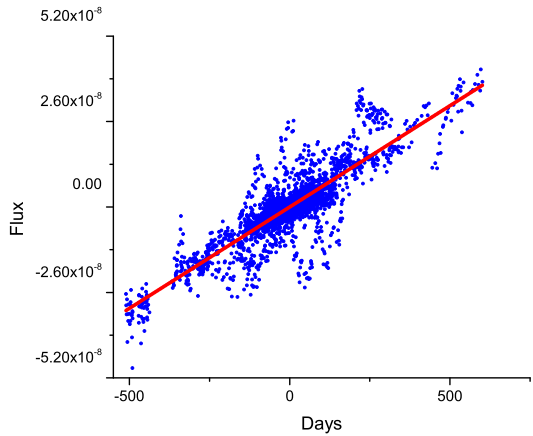}
\caption{The blue points are data points chosen from $13$ quasars: 6.7059.207;
13.5962.237;
9.4882.332;
11.8988.1350;
208.16034.100;
207.16316.446;
68.10972.36;
75.13376.66;
5.4892.1971;
59.6398.185;
13.5717.178;
13.6805.324;
1.4537.1642. The time and flux are shifted so that each segment is centered at the origin ($t=0$,$F=0$). The red line is the least-squares fit for the data. The slope is $k=6.47 \pm 0.07\times 10^{-11}$ units/day.}
    \label{slop}
\end{figure}
We focus on data with consistent observations over an interval longer than $90$ days,
 since that allows for more reliable identification (or rejection) of the observed patterns.
 We pick several straight line light curve segments from $13$ quasars and shift each segment
 so that it is centered at the origin ($t=0$,$F=0$). If the slope of the segment is negative,
 we multiply it by $-1$, since we are interested only in the absolute value of the slope.
 We then fit a straight line through the data collected from the corresponding segments
 of all $13$ quasars and find the slope, (see Fig.~\ref{slop}).
 We assume that all data points have the same weight in order to avoid a single segment giving a dominant contribution.
 The slope $k$ from the least-squares fit is
\be \label{mslope}
k= 6.47 \pm 0.07\times 10^{-11} \ {\rm units/day}.
\ee
where the ``unit" is $F^0$. If this pattern of parallel slopes in quasar light curves is not a coincidence, it would indicate that one can use quasars as distance standards, e.g. like Type Ia supernovae.

{\bf Finding the redshift of a quasar from its light curve - Linear fit.~}
We now illustrate the procedure of determining the redshift of an unknown quasar from its light curve.
We choose a quasar that was not included in our sample of $13$ quasars.
We identify five segments from its light curve that appear straight and parallel to one another, as shown in Fig.~\ref{example}
(the number of segments is not crucial -- the more the better).
We take each of those segments, discarding the rest of the light curve,
and shift them so that each segment is centered at ($t=0$,$F=0$).
Note that the time axis is in observer time, not quasar rest-frame time
since the redshift is ``unknown".
We fit the data collected from all the segments (all belonging to the same quasar)
with a single straight line and find the slope, as shown in Fig.~\ref{fit}.
In this case, the slope is $1.94 \pm 0.16\times 10^{-11}$ units/day.
Comparing this slope with that in Eq.~({\ref{mslope}) for the $13$ quasars rest-frame light curves,
we can calculate the redshift of that particular quasar
\begin{equation}
z=\frac{6.47 \pm 0.07\times 10^{-11}}{1.94 \pm 0.16\times10^{-11}}-1=2.34 \pm 0.27
\end{equation}
Indeed, this value is consistent with the known redshift of that quasar, $z=2.32$.

The same procedure can be repeated for the slower variations (marked by the thick green lines in Fig.~\ref{normalize}).
The results are again consistent within the statistical errors, though the error bars on the slope are larger -- presumably because
of the ``noise'' from the fast variations as well as  from other less coherent variations.
\begin{figure}[h]
   \centering
\includegraphics[width=3.2in]{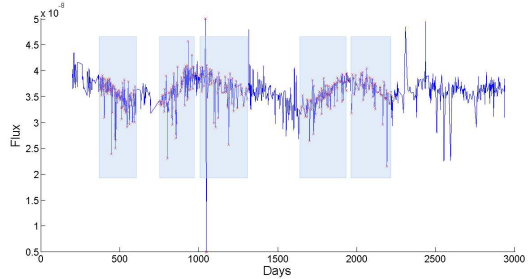}
\caption{The light curve from quasar $9.5484.258$ which was excluded from the step shown in Fig.~2. We choose for the linear fits those segments marked by shaded rectangles. }
    \label{example}
\end{figure}

\begin{figure}[h]
   \centering
\includegraphics[width=3.2in]{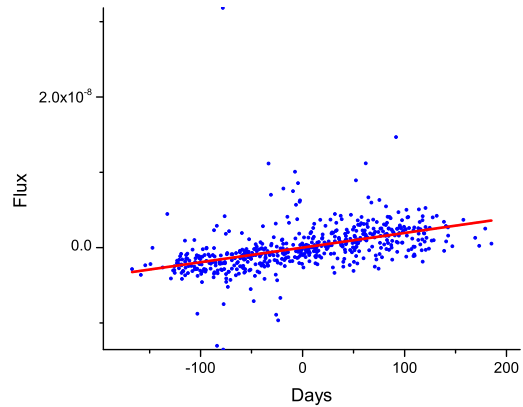}
\caption{Graph similar to Fig.~\ref{slop}, but only one quasar is included whose redshift we want to calculate. The best fit slope is $1.94\pm 0.16\times 10^{-11}$ units/day, which implies a redshift of $2.34\pm 0.27$. This agrees with the measured value $z=2.32$.}
    \label{fit}
\end{figure}
Therefore, if one can identify an oscillation mode that the particular segment of the light curve belongs to, one can readily calculate the redshift from the light curve.
Note that this implies that the time dilation (which is simply a counterpart of the redshift) is included in the quasar light curves, in contrast with conclusions in \cite{Hawkins:2010xg} .

{\bf Finding the redshift by matching the light curves.~}
Rather than identifying the segments through which we can fit a line as we did above,
we could  try to match the whole (or significant portions of the) quasar light curves.
We do not expect to match the curves from an arbitrary pair of quasars
as many different classes of quasars exist.
Fig.~\ref{12} shows the observed V-band light curves of two quasars of nearly identical redshifts:
{\it 206.17052.388} ($z=2.15$) and {\it 25.3712.72} ($z=2.17$).
The similarity is not apparent at all. However, for comparison, we can shift the time and flux origin of quasar {\it 25.3712.72} (and flip it):
\begin{equation}
\label{n}
t_n = t + 10 \quad  F_n = 6.3 \times 10^{-8}-F
\end{equation}
$t_n$ and $F_n$ are the time and flux after the shifts and flip.
\begin{figure}[h]
   \centering
\includegraphics[width=3.2in]{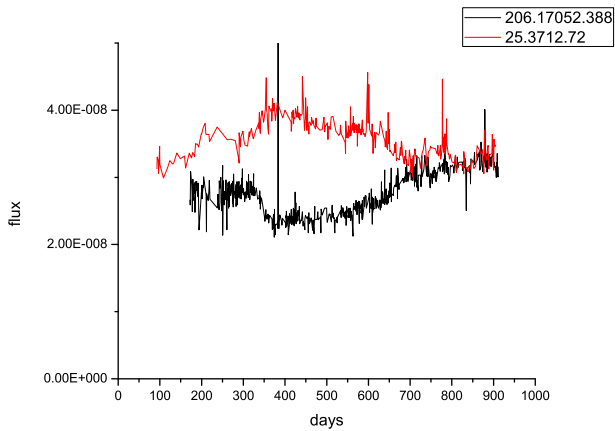}
\caption{
Light curves of two quasars. The black and red lines belong to quasars {\it 206.17052.388} and {\it 25.3712.72} respectively. The time has been normalized to the quasar's rest frame.}
    \label{12}
\end{figure}

\begin{figure}[h]
   \centering
\includegraphics[width=3.2in]{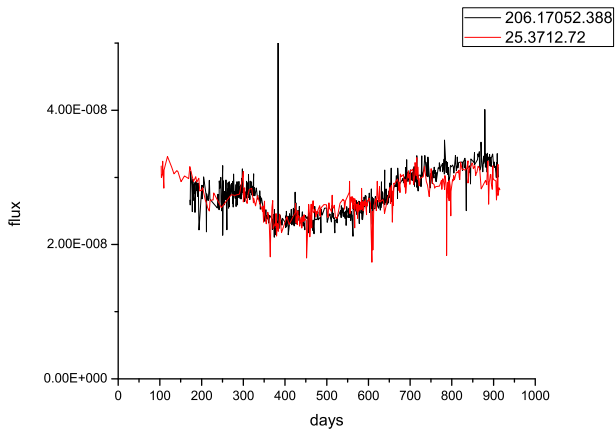}
\caption{The same quasars as in Fig.~\ref{12}, with the red one transformed according to Eq.~(\ref{n}). The match is now apparent.}
    \label{13}
\end{figure}
 Fig.~\ref{13} shows that the shifted light curves overlap with one another remarkably well.

We now define a procedure of finding the redshift of a quasar from its light curve.
To do this, we start with the observed quasar light curves,
for example {\it 11.8988.1350} ($z=0.33$) and {\it 5.4892.1971} ($z=1.58$)
We will keep the original data from {\it 11.8988.1350}, and manipulate the data from {\it 5.4892.1971} to obtain a match.
As in the previous examples, we will use four kinds of global transformations to shift the light curves.
The first one is the time shift ($\Delta t$) which just resets the initial time and has no special physical meaning.
The second one is a global offset of the flux intensity ($\Delta F$).
Our example in Figs.~\ref{12} and \ref{13} shows that $F_n(t)\rightarrow -F_n(t)$ is also a necessary transformation to allow.
Finally, we must rescale the time coordinate by a factor $\alpha_{12}= \frac{1+z_1}{1+z_2}$.
It is of course this factor $\alpha_{12}$ that we are most interested in obtaining.
Therefore, the complete set of allowed global transformations of the flux is
\begin{equation} \label{fn}
F_n(t)\rightarrow \pm F_n(\alpha_{12}(t-\Delta t) )-\Delta F
\end{equation}

Applying transformations in Eq.~(\ref{fn}) to {\it 5.4892.1971}, we can find a very good match with {\it 11.8988.1350}, as shown in Fig~\ref{ex2}. We also get an approximate value of $\alpha_{12}$.
Once we have matched the curves, we can identify the region where the light curves have the best overlap.
It is that region that we will use to fit the data statistically.
\begin{figure}[h]
   \centering
\includegraphics[width=3.2in]{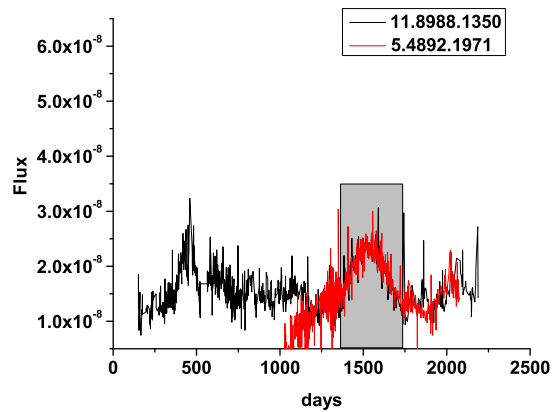}
\caption{The light curve of the quasar {\it 5.4892.1971} ($z=1.58$)  is transformed using only global transformation as in Eq.~(\ref{fn}) in order to match with {\it 11.8988.1350} ($z=0.33$). Matching the curves with significantly different redshifts perhaps eliminates contamination of space as the explanation of similarities. The gray rectangle represents the region of the greatest overlap which we will use for our fit.}
    \label{ex2}
\end{figure}

The main goal  is to find the value of $\alpha_{12}\equiv\frac{1+z_1}{1+z_2}$ which gives the best match to the data.
From Fig~\ref{ex2}, we identify the region where the light curves have the best match (gray area).
We extract the light curve data from these regions and transform one of them using $\Delta t$, $\Delta F$ and $\alpha$.
We then combine the data from these two light curves into a single light curve,
${\rm Flux}(\Delta t, \Delta F,\alpha_{12}; t_i )$, according to their new time sequence.

To check if a given transformation gives a good match, we  fit  ${\rm Flux}(\Delta t, \Delta F,\frac{1+z_1}{1+z_2};t_i )$
with a quartic polynomial:
\begin{equation}
S(t_i)\equiv a_4 t_i^4+a_3 t_i^3+a_2 t_i^2+a_1 t_i+a_0
\end{equation}

Fig.~\ref{fit2} shows the best fit, for the three-parameter ($\Delta t$, $\Delta F$ and $\alpha_{12}$) fit.
We have minimized
\be
\chi^2\equiv\sum_i ({\rm Flux}(t_i)-S(t_i))^2
\ee
with respect to $\Delta t$, $\Delta F$ and  $\alpha_{12}$.
 The standard deviation is $ \sigma^2\equiv\chi^2/(N-1)$,
 where $N$ is the total number of data points.
 Fig.~\ref{fit1} shows $\chi^2$ in units of $\sigma^2$ as a function of $\alpha_{12}$.
 (We have minimized with respect to $\Delta t$  and $\Delta F$.)
Within one $\sigma$ (or 68\% confidence level) $\alpha=0.508^{+0.012}_{-0.014}$.
This is  consistent with the actual measured value of $0.516$.

\begin{figure}[h]
   \centering
\includegraphics[width=3.2in]{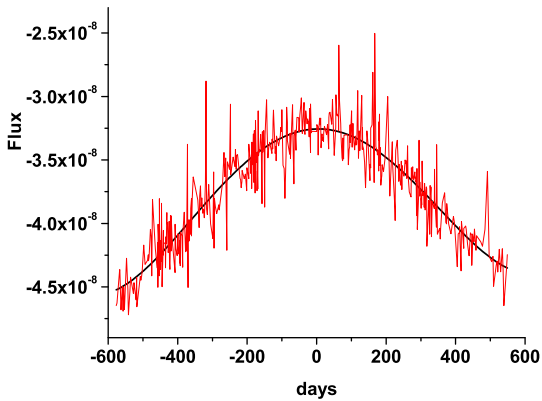}
\caption{The red line is the combined light curve ${\rm Flux}(\Delta t, \Delta F,\frac{1+z_1}{1+z_2},t_i )$ and the black line is the fitting function $S(t_i)$ (in our case quartic function)}
    \label{fit2}
\end{figure}

\begin{figure}[h]
   \centering
\includegraphics[width=3.2in]{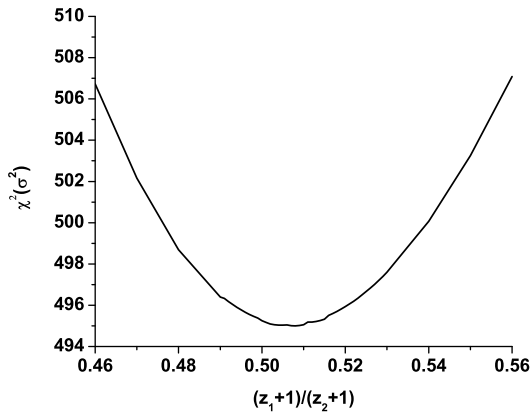}
\caption{$\chi^2$ in units of $\sigma^2$ for fitting ${\rm Flux}(\Delta t, \Delta F,\frac{1+z_1}{1+z_2},t_i )$ with the quartic function. $\chi^2$ has a minimum at the redshift ratio of $0.508$. The measured value is $0.516$.}
    \label{fit1}
\end{figure}

{\bf Conclusions.~}
By studying the data from MACHO quasars behind the Magellanic Clouds \cite{Geha:2002gv},
we observed patterns in quasar light curves that have previously gone unnoticed.
We analyzed the light-curves in two ways.
First, we characterized segments of the  light curves by the slopes of straight lines  through them.
These slopes appear to be directly related to the quasars' redshifts.
This allowed us to formulate a method for determining the redshift of an unknown quasar from its light curve.
The results match the known values extremely well.
This technique appears to allow us to obtain the redshifts of quasars with such linear trends within a few percent.
We also formulated an alternative method for determining the redshift that does not rely on a linear fit.
Matching the segments of two quasars light curves,
we were able to fit for the redshift ratios of the  quasars, again within a few percent.

These techniques suggest that similar patterns shared by
different quasars may allow them to be used as standard clocks (or candles) to quantify
luminosity distance.


We performed our analysis for V-band light curves, though a similar procedure could be carried out for other wavelengths.
We currently do not have a theoretical explanation of this effect.
We would not want to speculate much on the possible explanation since the physics of these objects is poorly understood.
Heuristically, if the frequency, $f$, and the corresponding amplitude, $A$, of the oscillation mode satisfy $fA=$constant,
and one looks at the sine-wave oscillations, then a constant slope $A\sin(ft)\sim Aft$ would appear for small $ft$.
Alternatively, it could happen that particular quasi-periodic quasar oscillations described in \cite{Chakrabarti:2004uu} (see also \cite{Lovegrove:2010te} where similar objects are studied) are behind this effect. For related studies see also \cite{MacLeod:2010qq,Kozlowski:2009my,Kelly:2009wy}. To identify the physics of the pattern we discovered clearly requires further investigation.

Regardless of its theoretical explanation, this observed effect  suggests that one might be able to use quasars as distance standards.
It will be important to extend the study to other quasars in order to find the dispersion of this effect.
This will require a larger sample  than the $14$ high quality quasar light curves, with continuous coverage over at least $90$ days,
that are currently available.

\begin{acknowledgments}
This work was partially supported by the US National Science Foundation, under Grants No. PHY-0914893 and PHY-1066278.
\end{acknowledgments}

\end{document}